\documentclass[aps,prb,twocolumn,superscriptaddress]{revtex4-2}

\usepackage{graphicx}
\usepackage{amssymb}
\usepackage{amsmath}
\usepackage{amsfonts}
\usepackage{braket,gensymb}
\usepackage{color}
\usepackage[utf8]{inputenc}		
\usepackage[english]{babel}
\usepackage{array}

\begin{document}
\title{
Magnon edge states of skyrmion crystal in non-uniform magnetic field 
} 
\author{V.~E. Timofeev}
\email{viktor.timofeev@spbu.ru}
\affiliation{St.Petersburg State University, 7/9 Universitetskaya nab., 199034
St.~Petersburg, Russia} 
\affiliation{NRC ``Kurchatov Institute'', Petersburg Nuclear Physics Institute, Gatchina
188300, Russia}
\author{D.~N. Aristov}
\affiliation{NRC ``Kurchatov Institute'', Petersburg Nuclear Physics Institute, Gatchina
188300, Russia}
\affiliation{St.Petersburg State University, 7/9 Universitetskaya nab., 199034
St.~Petersburg, Russia} 

\begin{abstract} 
A regular lattice of magnetic skyrmions is the ground state of 
thin ferromagnetic films with Dzyaloshinskii-Moriya interaction in a relatively wide range of external magnetic fields.  It was previously theoretically shown that upon the increase of magnetic field a topological transition in the magnon spectrum of such skyrmion crystal (SkX)  may occur. Non-uniform magnetic field may lead to localized magnon states emerging at the interface between two half-planes of SkX. 
 Using semiclassical quantization and the stereographic projection approach, we study such appearing edge states both in a full band structure calculation and in simplified effective model. The latter effective model described by extended Dirac equation  is applicable to two relevant magnon bands near $\Gamma$ point. We show that both the chirality of emerging edge states and the degree of its localization at the interface is controlled by magnetic field profile. We demonstrate that the localization length may be as small as a few inter-skyrmion distances. 
 
\end{abstract}

\maketitle

\section{Introduction}

Magnetic skyrmions are whirls of local magnetization characterized by a nontrivial topological charge. In chiral magnets with Dzyalosinskii-Moriya interaction (DMI) \cite{bogdanov1989thermodynamically,Bocdanov_1994} interplay of ferromagnetic exchange, DMI and external magnetic field stabilizes skyrmions and arranged them into regular lattices, known as skyrmion crystal (SkX)\cite{Nagaosa2013}. In bulk samples SkX is stable in a small region of the phase diagram near critical temperature \cite{muhlbauer2009skyrmion}. Region of SkX stability widely extends in temperature-field phase diagram in thin films, including temperatures near $T=0$\cite{Tonomura_2012, Seki_2012}.

Non-collinear structure of the local magnetization induces an emergent gauge field in the equation of motion for elementary excitations \cite{schutte2014magnon}. This gauge potential, which acts as an effective vector potential for magnons, gives rise to a rich band structure. The spectrum of excitations contains topologically trivial and nontrivial magnon modes \cite{garst2017collective, Timofeev2022}. The low-energy excitations of the SkX can be interpreted as  deformations modes: displacements \cite{petrova2011,Timofeev_2023b}, elliptic distorions, dilatation or breathing (Br) modes, clockwise (CW) and counterclockwise (CCW) rotations \cite{garst2017collective, Timofeev2022} etc. Due to selection rules only Br, CW and CCW modes manifest itself in magnetic resonanse experiments \cite{mochizuki2012, onose2012observation, Schwarze2015}, while other branches require hybridization via symmetry-breaking perturbations to be detected\cite{takagi2021} in that kind of experiments. These and other types of excitations were also observed in inelastic neutron scattering\cite{Weber_2022}.

The principle of bulk-boundary correspondence states that nontrivial topology of a bulk band structure leads to appearance of edge states in a finite samples. This principle works not only for electron systems, but for magnons too \cite{McClarty_2022}. Due to the technological applications into magnonics, magnon topological insulators were recently widely discussed, see \cite{Zhang_2013,Zhu_2021}. It was shown that stabilization of edge magnon states may be provided for some systems with DMI by interaction between magnons  \cite{Mook_2021}. In this context SkX seems to be a natural platform for studies of chiral topological magnons and magnon edge modes.

It was particularly shown that nontrivial topological character of SkX band structure may by itself lead to appearance of edge states in the samples with finite geometry \cite{roldan2016}. It was shown there that if 
the external magnetic field is directed in opposite way in two SkX domains, then 
the localized edge magnon mode appears on the interface between these two domains. 
The method of micro-magnetic simulation, using generalized Thiele equation,  shows the existence of chiral and non-chiral edge modes  at the edge of finite honeycomb lattice of skyrmions \cite{PhysRevB.98.180407}.

The SkX band structure is highly tunable, depending on different factors such as external magnetic field\cite{Timofeev2022}, dipole-dipole interactions\cite{garst2017collective} etc. Magnetic field is perhaps the simplest parameter to tune in laboratory. Notably, a topological transition occurs in the minimal SkX model (a 2D ferromagnet with DMI and an external field) as the field strength increases: the gap between the CCW and Br modes closes, accompanied by a redistribution of Berry curvature and changes in the Chern numbers of the bands \cite{Diaz2020,Timofeev_2023c}. Authors of Ref.\ \cite{Diaz2020} discuss a possibility to use magnetic field as a control parameter to switch on and off the  magnon spin current carried by topologically protected chiral magnonic edge states. 
 To the best of our knowledge, such topological transition in the excitation spectrum of SkX was not yet experimentally observed, although the relevant gap reduction with the increase of magnetic field has been reported in \cite{takagi2021}.
In the present work we study edge magnon modes appearing on an interface of two half-planes with different magnitude of external magnetic field, pointing in the same direction. We assume that this is the simplest configuration of systems where edge magnons can appear and be observed.

The rest of the paper is organized as follows.  In Section \ref{sec:model} we introduce our model and briefly discuss methods and techniques that we will need in further calculations. In Section \ref{sec:toptrans} we consider a reduced ``toy'' model with two branches of excitations and exhibiting  topological transition. Here we find an exact solution in a situation, when parameters of the reduced Hamiltonian vary in space, and we discuss an appearance of excitations localized at the interface of two domains with different topological properties of the spectrum. Section \ref{sec:numerical} is dedicated to full analysis of excitation spectrum of SkX in a geometry of elongated stripe, we also discuss obtained results. 
The localized character of the edge state is analyzed in Section \ref{sec:IPR}. 
The concluding Section \ref{sec:conc}   summarizes the main results of our work.

\section{Dynamics of skyrmion crystal}
\label{sec:model}


\subsection{Model of ferromagnetic film}

We study a continuous model of a thin ferromagnetic film with Dzyaloshinskii-Moriya interaction (DMI) in external magnetic field, with the following energy density:
\begin{equation}
\mathcal{E} =   \frac{C}{2}  (\nabla \mathbf{S})^2 + 
D(\mathbf{S},\nabla\times\mathbf{S})  - B  S_{z},
\label{eq:classicalenergy}
\end{equation}
where $C$ is ferromagnetic stiffness, $D$ is DMI constant, and $B$ is proportional to external magnetic field. For simplicity we ignore demagnetization field, dipole-dipole interaction, anisotropy terms etc.  We also assume that magnetization doesn't change across the thickness of the film. In the low temperature limit the local magnetization value is uniform and saturated, $|\mathbf{S}| = S$.

\begin{figure}[t]
\center{\includegraphics[width=0.99\linewidth]{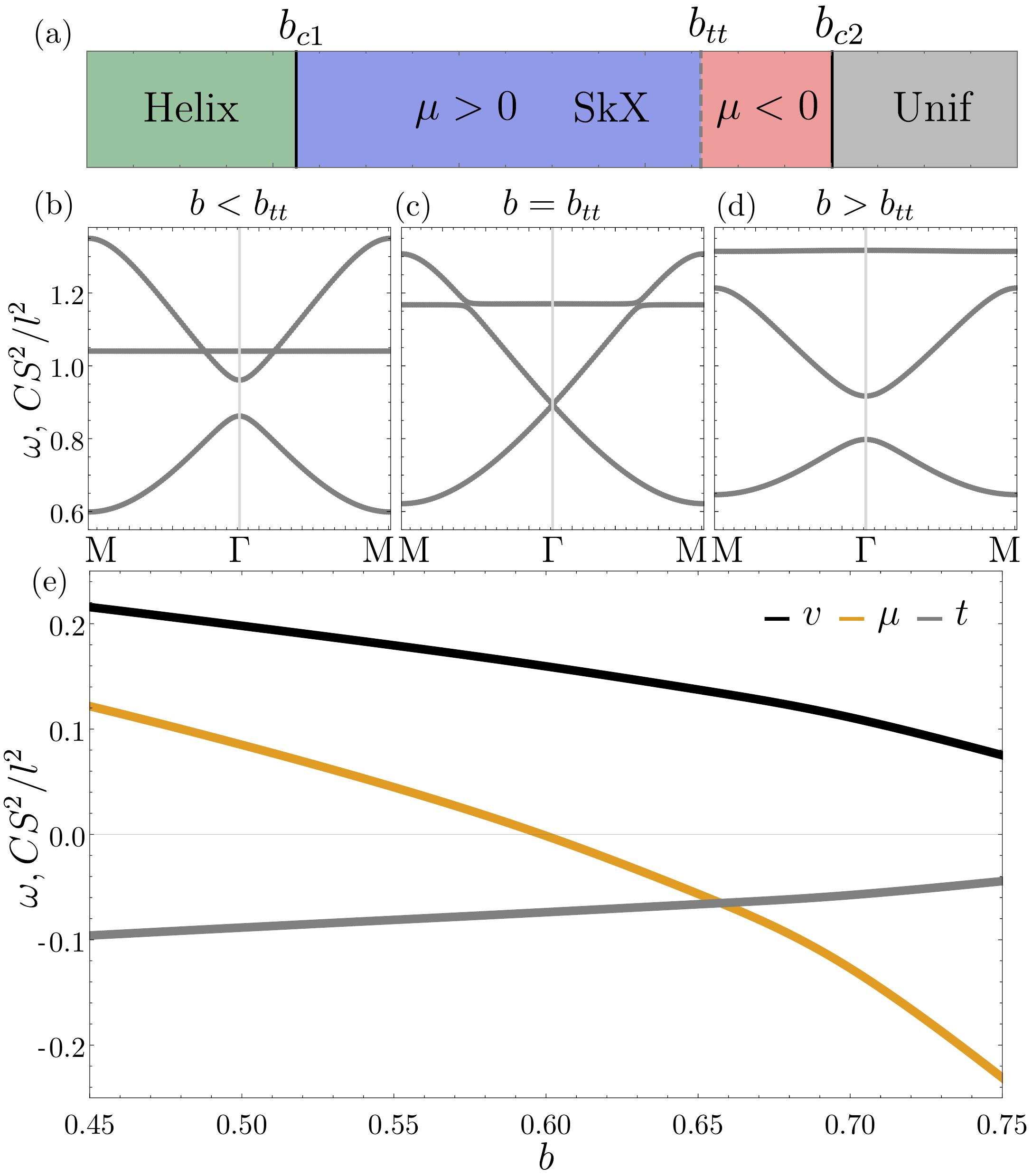}}
\caption{\label{fig:phases}
Sketch (a) of a phases shows order of a different magnetic phases in a model \eqref{eq:classicalenergy2}: helix, SkX and uniform configuration. Panels (b), (c) and (d) show evolution of Br and CCW branches with increasing of external magnetic field $b$. (e) dependencies parameters of Hamiltonian \eqref{eq:redham} of magnetic field $b$.}
\end{figure}

It is convenient to define a characteristic length, $l=C/D$, and measure distances in units of $l$. Similarly, the energy density is  measured in units of $ CS^2 l^{-2} = S^2D^2/C$.  The model \eqref{eq:classicalenergy} depends in  these units only on one dimensionless parameter $b=BC /SD^2$ and has the form : 

\begin{equation}
\mathcal{E} =   \tfrac{1}{2}  (\nabla \mathbf{n})^2 + 
(\mathbf{n},\nabla\times\mathbf{n})  - b \, n_{z}\,,
\label{eq:classicalenergy2}
\end{equation}
where $\mathbf{n}=\mathbf{S}/|\mathbf{S}|$, and we assume that $b>0$. It is well known that the model \eqref{eq:classicalenergy2} has three different ground states (see Fig.\ref{fig:phases} (a)): magnetic helix at $b<b_{c1}\approx0.225$, SkX at $b_{c1}<b<b_{c2}\approx0.8$, and induced ferromagnetic state at $b>b_{c2}$, when the local magnetization becomes parallel to magnetic field.

\subsection{Stereographic projection approach}

We use the stereographic projection approach in discussion of skyrmion configurations. The unit vector along the local magnetization, $\mathbf{n}$,  is expressed in terms of complex-valued function, $f$, and its  conjugate, $\bar{f}$ :
\begin{equation}
n_1 + i \, n_2  = \frac{2f}{1 + f\bar{f}}\,,\quad 
n_z = \frac{1 - f\bar{f}}{1 + f\bar{f}}\,.
\label{eq:stereo}
\end{equation}
This definition of stereographic function \eqref{eq:stereo} means that zeros of function $f$ correspond to $n_z=1$, i.e ''up`` direction of magnetization, and poles of $f$ correspond to opposite direction. 
A minimum of the  energy \eqref{eq:classicalenergy} is provided by an appropriate  choice of $f$. 
In particular, the induced ferromagnetic state at $b>b_{c2}$ corresponds to $f\equiv0$. 

 We showed previously \cite{Timofeev2019} that the stereographic function of SkX can be represented, with a good accuracy, by a sum of stereographic functions of individual skyrmions centered at positions $\mathbf{r}_i$: 
\begin{equation}
    f_{SkX} = \sum_i f_1 (\mathbf{r}-\mathbf{r}_i)\,, 
  \label{SkX0}  
\end{equation} 
due to localized character of single skyrmions in external magnetic field and the repulsion between them. The trial function $f_{1}(\mathbf{r})$ of one skyrmion for the model \eqref{eq:classicalenergy} can be chosen  in the absence of magnetic anisotropy in a following form:
\begin{equation}
f_1(z,\bar{z})=i\frac{\, z_0 \,\exp{(- \alpha |\bar{z}|^2/z_0^2)}}{\bar{z}}\,,
\label{eq:anz}
\end{equation}
with $z=x + iy$ and $\bar{z}=x - iy$, here $z_0$ is a real number associated with a size of skyrmion, $\alpha$ is a dimensionless parameter defining a decay of skyrmion profile at large distances. The imaginary unit prefactor,  $i$,  in \eqref{eq:anz} means that we consider Bloch-type skyrmions, which minimizes the energy of DMI term in the model \eqref{eq:classicalenergy}.

Let us note that our trial function \eqref{eq:anz} is not unique. We can alternatively use the widely used cylindrical magnetic domain (CMD) or magnetic bubble ansatz for individual skyrmion, whose stereographic projection has the form ~:
\begin{equation}
f_{\text{CMD}}(z,\bar{z}) = i\frac{|\bar{z}|\sinh{R/\delta}}{\bar{z} \sinh{|\bar{z}|/\delta}}\,,
\label{eq:360dw}
\end{equation}
where $R$ is skyrmion radius, and $\delta$ is a domain wall width. Each ansatz, \eqref{eq:anz} and \eqref{eq:360dw},  has two variatonal parameters, and they give quite similar optimal static configuration of SkX.


\subsection{Semiclassical magnetization dynamics} 

In this subsection we discuss the Lagrangian formalism for semiclassical magnetization dynamics, which is equivalent in  lowest order to conventional linear spin-wave theory of magnetically ordered systems. 
Non-dissipative dynamics of local magnetization is usually described in terms Landau-Lifshitz (LL) equation: $\dot{\mathbf{S}}=-\gamma_0 \, \mathbf{S}\times\mathbf{H}$, where $\gamma_0$ is a gyromagnetic ratio and $\mathbf{H}=\delta {E}/\delta\mathbf{S}$ is an effective magnetic field. The LL equation could be considered as an Euler-Lagrange equation for Lagrangian with correspondent kinetic term \cite{Doering1948}. In terms of stereographic projection \eqref{eq:stereo} could be written in the following form\cite{metlov13vortex}:
\begin{equation}
\mathcal{T}=  \frac i2 \frac{\bar f \partial_t f - f \partial _t \bar f}
{1+f \bar f}\,.
\label{kinLagrangian}
\end{equation}
with $\gamma_0$ absorbed into the definition of time, $t$.  

The precise consideration of a LL equation for non-collinear magnets is hardly available, and we consider the low-energy dynamics of local magnetization   using semi-classical approach. We consider small time-dependent fluctuations about the static ground state configuration: 
\begin{equation}
f(\mathbf{r},t) = f_0(\mathbf{r}) + (1 + f_0(\mathbf{r}) \bar{f}_0(\mathbf{r}))\psi(\mathbf{r},t)/\sqrt{2S}\,.
\end{equation}
With this assumption the Lagranian density $\mathcal{L}=\mathcal{T} - \mathcal{E}$ takes form of formal $1/\sqrt{2S}$ expansion: $\mathcal{L} = \mathcal{L}_0 +   \mathcal{L}_1 +   \mathcal{L}_2 + \ldots$. 
In this expansion  $\mathcal{L}_0$ is time independent, $\mathcal{L}_1$ vanishes when  $f_0(\mathbf{r})$ delivers an energy minimum. 
The quadratic terms in $\psi$ corresponds to $\mathcal{L}_2$ and, in turn, to the linear spin-wave theory. We do not consider higher order terms, $\mathcal{L}_j$ with $j>2$, which would describe the interaction of spin waves.

The quadratic part of Lagrangian $\mathcal{L}_2$ takes the following form:
\begin{equation}
\mathcal{L}_2 =\frac{1}{2} 
\begin{pmatrix}
  \bar{\psi},& \psi
\end{pmatrix}
\left(i
\begin{pmatrix}
  \partial_t& 0\\
  0& -\partial_t
\end{pmatrix}
-\hat{\mathcal{H}} \right)
\begin{pmatrix}
  \psi\\
  \bar{\psi}
\end{pmatrix},
\label{Lagr2}
\end{equation}
with the Hamiltonian operator 
$\hat{\mathcal{H}}$ of the form 
\begin{equation}
\hat{\mathcal{H}}=
\begin{pmatrix}
  (-i\nabla + \mathbf{A})^2 + U&
   V\\
  V^*&
   (i\nabla + \mathbf{A})^2 + U
\end{pmatrix} \,,
\label{ham}
\end{equation}
with $\nabla =  \mathbf{e}_x \partial_x + \mathbf{e}_y \partial_y$. 
Here $U$, $V$ and $\mathbf{A} = \mathbf{e}_x A_x + \mathbf{e}_y A_y$ are rather cumbersome  functions of the static function $f_0(\mathbf{r})$ and its gradients, listed in  \cite{Timofeev2022}.

The Lagrangian \eqref{Lagr2}  results in 
the Euler-Lagrange equation
\begin{equation}
-i \frac{d}{dt}
\begin{pmatrix}
  \psi\\
  \bar{\psi}
\end{pmatrix}
=\sigma_3\hat{\mathcal{H}}
\begin{pmatrix}
  \psi\\
  \bar{\psi}
\end{pmatrix},
\label{eq:BdGeq}
\end{equation}
with the energies of normal modes, $ \epsilon_n$ defined by  equation:
\begin{equation}
\Big( \epsilon_n \, \sigma_3 - \hat{\mathcal{H}} \Big) 
\begin{pmatrix}
  u_n\\
  v_n
\end{pmatrix}  
=  0 \,, 
\label{eq:shr}
\end{equation}
with $\sigma_3$ the   Pauli matrix. Notice that the characteristic equation \eqref{eq:shr} for $\epsilon_n$ corresponds to Bogoliubov coefficients, 
$ u_n$,  $v_n$, satisfying orthogonality relation 
 $ \int d\mathbf{r} \,( u_{n}^{*}  u_{m}  - v_{n}^{*}  v_{m}  )  = \delta_{nm}$, 
see \cite{Timofeev2022} for more details.

\subsection{Topological transition in the spectrum}

The solutions of Eq. \eqref{eq:shr} for  SkX in a uniform magnetic field were thouroughly analyzed in \cite{Timofeev2022}. The band structure of elementary excitations   is quite complicated and contains both topologically trivial bands with vanishing Berry curvature throughout the entire first Brillouin zone, and topologically nontrivial bands characterized by non-zero Chern numbers.

It was also shown that there is a topological transition in the spectrum of excitations of SkX \cite{Diaz2020, Timofeev_2023c}.   
At moderate values of the field, $b\le 0.6$,  two lowest topologically nontrivial branches are counterclockwise (CCW) and breathing (Br) modes, separated by the gap  at the center of the first Brillouin zone, $q=0$. 
Upon the increase of the field,  the energy gap between CCW and Br modes closes at the value  $b=b_{tt}\approx0.66$. 
The further increase of the field re-opens the gap, but the integral topological weight (Chern number) vanishes. 

In order to qualitatively analyze the properties of the topological transition, we extracted in \cite{Timofeev_2023b} a part of the full Hamiltonian 
which i) corresponds to CCW and Br magnon branches and ii) becomes degenerate at $b=b_{tt}$. We call it the reduced Hamiltonian and consider it in the next section. 
 

\section{Topological transition 
in the reduced model}
\label{sec:toptrans}

\subsection{Reduced Hamiltonian for CCW and Br modes}

\begin{figure*}[t!]
\center{\includegraphics[width=0.99\linewidth]{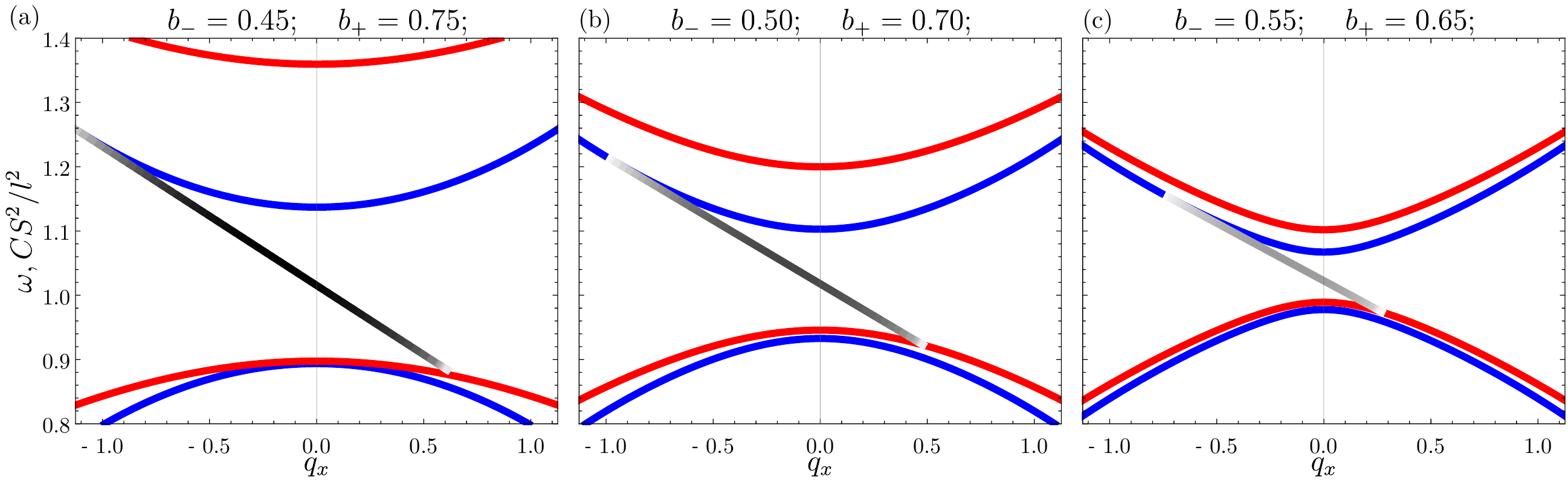}}
\caption{ \label{fig:params}
(a)-(c) are dispersion of excitation for reduced model \eqref{eq:disp} for different bottom and top magnetic fields, gray lines show interface mode dispersion tha appears on the interface, gradient from white to black corresponds to IPR value of wave functions, see Eq.\eqref{eq:ipr1}.}
\end{figure*}


When constructing the desired reduced Hamiltonian for  CCW and Br modes, we use the approach to description of 
the Goldstone (gyrotropic) mode of the skyrmion crystal.
The latter mode corresponds to displacements  of skyrmion centers, $\mathbf{u}$, out of their equilibrium position $\mathbf{r}^{(0)}$.  
We have for the stereographic function of perturbed SkX ~: 
%
\[
\sum_i f_1 (\mathbf{r}-\mathbf{r}^{(0)}_i - \mathbf{u}_i)  \simeq f_{0}(\mathbf{r}) -  \sum_{i} \mathbf{u}_i \nabla f_1 (\mathbf{r} - \mathbf{r}^{(0)}_i)\,,
\]
where $\mathbf{u}_i$ is a displacement of i-th skyrmion. The total perturbation is a sum of contributions from single skyrmions, each one can be represented as $ {u}_i \partial  f_1 (\mathbf{r} - \mathbf{r}^{(0)}_i) /\partial z + 
\bar{u}_i \partial  f_1 (\mathbf{r} - \mathbf{r}^{(0)}_i) /\partial \bar{z} $ with complex-valued amplitudes, $u_{i}= u^{x}_{i}+iu^{y}_{i}$, etc. 
After some calculations one can find the dispersion and Green's function for the 
gyrotropic mode, see details in \cite{Timofeev_2023b}.

By analogy with the gyrotropic mode, the Br and CCW modes are described as a sum of deformations of individual skyrmions, 
possessing 
the symmetry of the specific mode, namely:
\begin{equation}
\begin{aligned}
\mbox{Br:} &\quad  
u_{br} \, z\, \partial  f_1(\mathbf{r})/\partial z + \bar{u}_{br}\, \bar{z} \,\partial  f_1(\mathbf{r})/\partial \bar{z}\,, 
\\
\mbox{CCW:}&\quad  
u_{ccw} \, z^2 \partial  f_1(r)/\partial z + \bar{u}_{ccw}\,\bar{z}^2 \partial  f_1(r)/\partial \bar{z}\,, 
\end{aligned}
\end{equation}
where $u_{br}$ and $u_{ccw}$ are complex valued amplitude of each skyrmion's deformation, depending on a site in the lattice.

Performing Fourier transform of {  breathing} and {  CCW } modes, $ u_{j}  = \sum _\mathbf{q} e^{i \mathbf{q} \mathbf{R}_j} u_{\mathbf{q}}$, we 
obtain   the reduced Lagrangian as a  quadratic form in   ``spinor'': $\Psi^\dagger_ {\mathbf{q}} = \begin{pmatrix} 
\bar {u}_{br,\mathbf{q}} ,&  u_{br,\mathbf{q}}   ,&   \bar {u}_{ccw,\mathbf{q}} ,&  u_{ccw,\mathbf{q}} 
\end{pmatrix}$. After a series of algebraic manipulations, including unitary transformations, etc., we 
restrict our consideration to  two positive-energy branches of the spectrum, and reduce all dynamics to $2\times2$ Hamiltonian near $q=0$, see \cite{Timofeev_2023c} for details. 


Proceeding this way, we showed  
 that Br and CCW branches can be described near $\Gamma$ point, $q=0$,  by the following reduced 2$\times$2 Hamiltonian:
\begin{equation}     
\begin{aligned}  
\mathcal{H}_{2} & = 
	 \begin{pmatrix}
	E_0 + \mu + t q^2 , & v(q_x-iq_y) \\  v(q_x+iq_y) , &E_0 -\mu - t q^2
	\end{pmatrix}\,,
\end{aligned}
\label{eq:redham}
\end{equation}
here parameters $E_0,\mu,t$ and $v$ depending on external magnetic field, $b$, and $q^2=q_x^2 + q_y^2$.
By construction, $q$ is measured in inverse inter-skyrmion distances, $d^{-1}$. \cite{Timofeev_2023c}
Expression \eqref{eq:redham} is a well known 2$\times$2 Dirac Hamiltonian with additional quadratic term. The new ``spinor'' roughly corresponds  to 
$\Psi^\dagger_ {2,\mathbf{q}} = \begin{pmatrix} 
\bar {u}_{br,\mathbf{q}} ,  ,&   \bar {u}_{ccw,\mathbf{q}} 
\end{pmatrix}$.

The energies of the Hamiltonian \eqref{eq:redham} are found from equation, $(\mathcal{H}_{2}- \epsilon)\psi =0$, and are  given by:
\begin{equation} 
	\epsilon_{\pm,\mathbf{q} } = E_0 \pm \sqrt{ (\mu + t q^2) ^2+v^2q^2 }\,.
\label{eq:disp}
\end{equation}
One  observes that the gap between the two bands closes, $\epsilon_{+,0}-\epsilon_{-,0}=0$, at  $\mu=0$. 
The eventual change of the sign of $\mu$ leads to re-opening of the gap.
This transition in the spectrum is accompanied by a  change in Chern numbers of branches: $C_{\pm} = \pm(\mbox{sign}\,t  -  \mbox{sign}\,\mu)/2$. In the simplified model \eqref{eq:redham} we have $b_{tt}=0.6$, see Fig.\ref{fig:phases}(e), it slightly differs from value of transition field in full model.
The bulk-boundary correspondence principle states that the change of Chern numbers of bands leads to the change in number of edge states (by analogy with topological insulators). 

 The magnon edge states in finite size systems were investigated in Ref.\ \cite{Diaz2020}. It was shown there that for $b<b_{tt}$ edge states are topologically protected against disorder, and they become vulnerable to boundary disorder at fields above $b_{tt}$.



\subsection{Domain wall and the interface mode}

 Let us now consider two half-planes with different magnitude of $b$. We assume that the  magnetic field is nonuniform:
\begin{equation}
b(y)=
    \begin{cases}
     b_{+},\, y>0\\
     b_{-}, \, y<0
    \end{cases}\,,
\label{eq:nonunifield}
\end{equation}
with $b_{+}>b_{tt}$ and $b_{-}<b_{tt}$, i.e. the parameters of Hamiltonian \eqref{eq:redham} exhibit a jump discontinuity on the interface. This situation is  usually called {\it domain wall}, and there is an analytical solution for this configuration of a field. 
We sketch the corresponding derivation below. 

We consider a half-plane with uniform magnetic field and seek an eigenfunction of the Hamiltonian \eqref{eq:redham} with the well-defined wave-vector  along the edge of the half-plane:
\begin{equation}
\Psi_{q_x}(x,y)=\exp(iq_xx)\psi(y)\chi,
\label{eq:half-planesol}    
\end{equation}
where $\chi$ is a vector and $\psi(y)$ is a scalar function of spatial coordinate, $y$. Representing   $q_y$ in \eqref{eq:redham} as differential operator, $q_y = -i\partial_y$, one gets a set of second-order differential equations $(\mathcal{H}_{2}-\epsilon(q_x)I)\chi\psi(y)=0$, with a solution: $\psi(y) \propto \exp(\lambda y)$. We consider a characteristic polynomial, $\mbox{det}|\mathcal{H}_{2}-\epsilon(q_x)I|=0$, as a biquadratic equation on $\lambda$, with four solutions, two of them positive and two of them  negative.

\begin{figure}[t!]
\center{\includegraphics[width=0.99\linewidth]{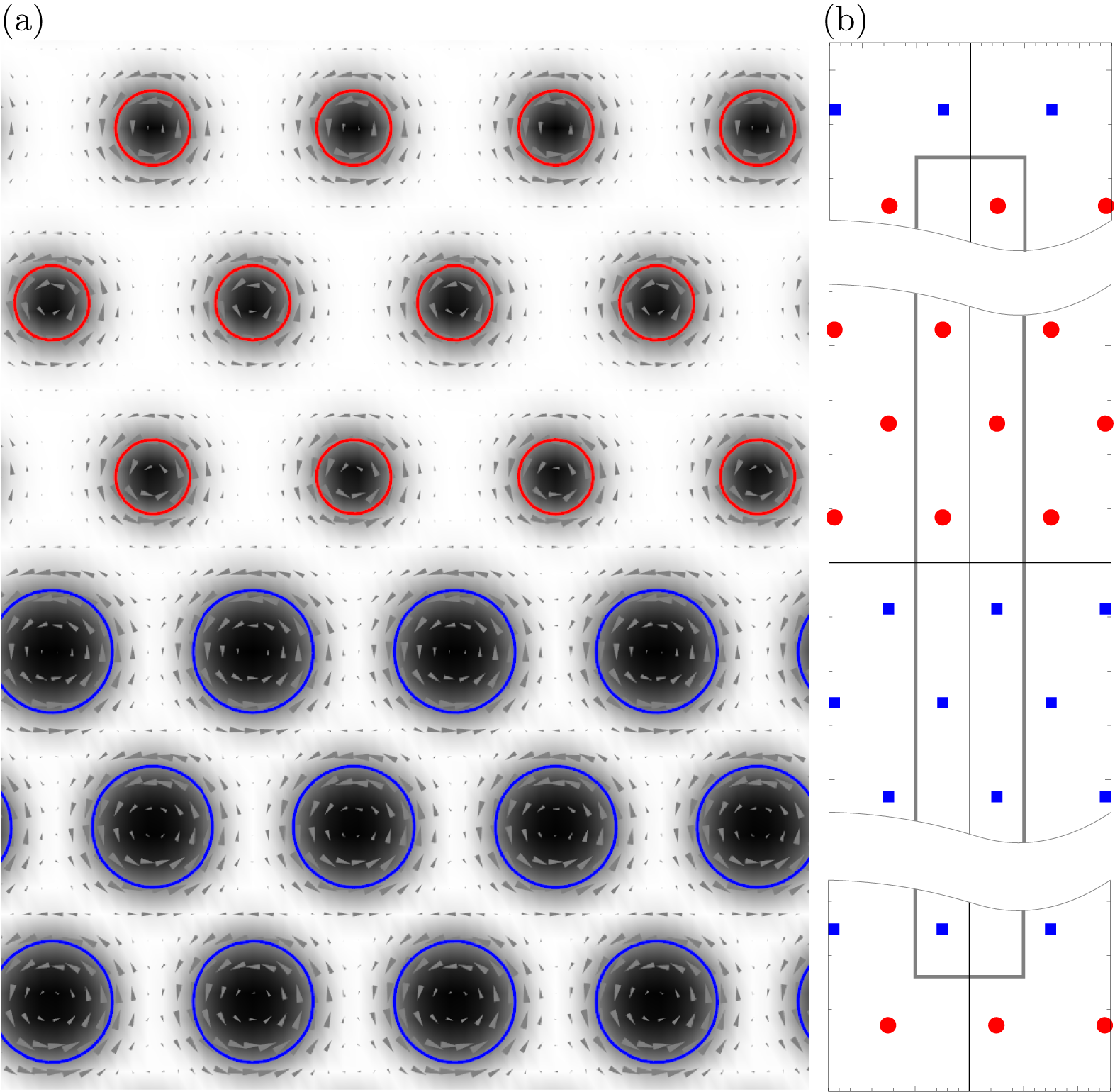}} 
\caption{ \label{fig:statconfig}  
(a)  Skyrmion configuration for $b_{+}=0.76$ and $b_{-}=0.26$, gray arrows shown in-plane component of local magnetization, shades of gray demonstrate value $n_z$, where white corresponds to $n_z=1$, and black to $n_z=-1$. Red and blue circles demonstrate skyrmion radii. (b) Red circles and blue dots correspond to position of small and large skyrmions, gray line demonstate boundaries of primitive stripe-cell.}
\end{figure}

We are interested in the solution, localized at the  interface, so we pick two positive lambdas, $\lambda>0$, for the half-plane with $y<0$ and $\lambda<0$ for $y>0$. 
Requiring continuity of $\psi(y)$ and its  first derivative at the interface, we write:
$(\Psi_{-} - \Psi_{+})\big|_{y=0}=0$ and $(\partial_y\Psi_{-} - \partial_y\Psi_{+})\big|_{y=0}=0$. These equations  define the dispersion relation for the localized mode, $\epsilon(q_x)$. Such a solution exists only for areas with different band topology, i.e. $\mbox{sign}(\mu_{+}t_{+})\neq\mbox{sign}(\mu_{-}t_{-})$.
  The direction of group velocity $\mathbf{v}=\nabla_{\mathbf{q}}\epsilon $ depends on external magnetic field, which is directed in our case along the unit vector, $  \mathbf{n}_{b} $, normal to the plane. Let us define  $ \mathbf{n}_{int} $ as an in-plane unit vector perpendicular to the interface and directed from topologically non-trivial area,  $b_{-}$, to topologically trivial one, $b_{+}$.  Then we have $\mathbf{v}= | \mathbf{v}| \, \mathbf{n}_{b}\times\mathbf{n}_{int}$, and the group velocity for our configuration in Fig.\ \ref{fig:params}   is directed along $x$ axis with a negative sign. 
The degree of localization of this interface mode is conveniently characterized by inverse participation ratio (IPR) criterion, and we postpone this discussion until Sec. \ref{sec:IPR} below.  


Having considered the interface mode in our simplified toy model, we perform now a calculation in full model, characterized by Eq. \eqref{eq:BdGeq}. 

\section{full model calculation}
\label{sec:numerical}

\begin{figure*}[t]
\center{\includegraphics[width=0.99\linewidth]{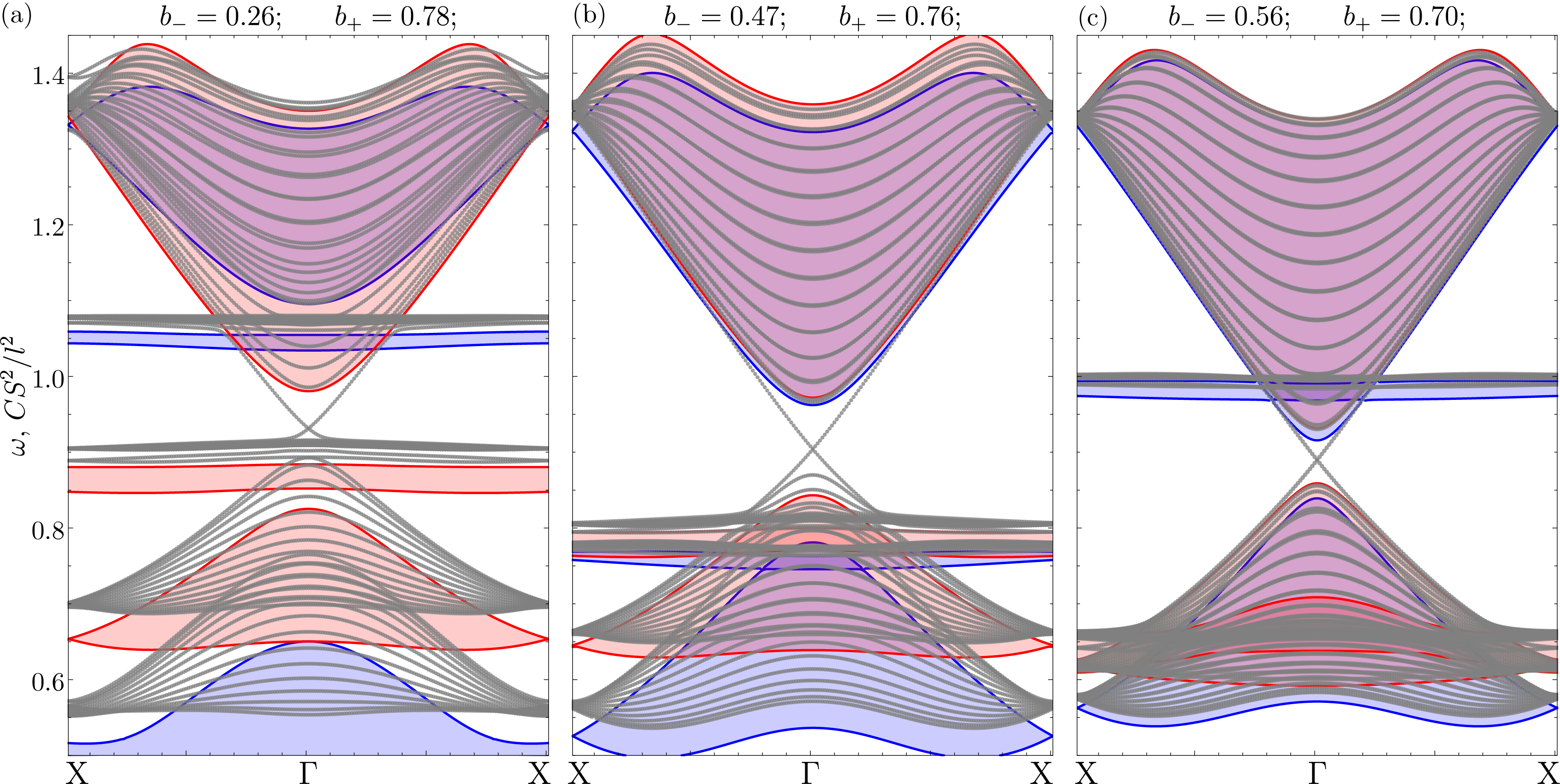}}
\caption{\label{fig:dispers}
Gray dots on panels (a)-(c) are dispersions of elementary excitations in the stripes under different values of external magnetic fields. Red and blue areas show dispersion of normal modes of uniform SkX under $b_{+}$ and $b_{-}$ respectively.
}
\end{figure*}

\subsection{Skyrmion configuration under nonuniform magnetic field}

We consider two half-planes, with different values of the external magnetic field \eqref{eq:nonunifield}. 
The stereographic function of this configuration is written as, cf.\ Eq.\ \eqref{SkX0}:
\begin{equation}
    f = \sum\limits_{y_i<0} f_1(\mathbf{\mathbf{r} - \mathbf{r}}_i,z_0,\alpha) + \sum\limits_{y_i>0} f_1(\mathbf{r} - \mathbf{r}_i,z_0',\alpha')\,,
\label{eq:stereoSkX_halfplanes}
\end{equation}
where $\mathbf{r}_i = n_i \mathbf{a}_1 + m_i \mathbf{a}_2$ is a lattice translation, with $\mathbf{a}_1 = d\, \mathbf{e}_x$ and $\mathbf{a}_2 = \tfrac12 d\, \mathbf{e}_x   +  \tfrac{\sqrt{3}}{2}d\,\mathbf{e}_y  $. Thus the energy of this configuration depends on five parameters: $d$, $z_0$, $z_0'$, $\alpha$ and $\alpha'$.

Notice that  Eq. \eqref{eq:stereoSkX_halfplanes}  assumes that non-uniform magnetic field affects only skyrmion size, $z_0$, and profile function, $\alpha$, but the distance between neighboring skyrmions, $d$, is intact. This assumption is in accordance with experimental observations and also with our modeling in the considered range of magnetic field. \cite{Timofeev2021}

A primitive cell of this configuration is an infinitely long stripe extended along the $y$-axis, with the width, $d$, along the $x$-axis. For simplicity of numerical analysis we consider a finite size stripe, see Fig.\ \ref{fig:statconfig}(b), which contains $N_{sk}$ skyrmions in each of   the  half-planes.  
The length of this rectangular stripe-cell is  thus $\sqrt{3} d N_{Sk}$.  
The periodical condition along the $y$-axis is imposed,  as shown in Fig.\ \ref{fig:statconfig}(b), by adding additional skyrmions of other 
kind beyond the length of the stripe.  
The energy \eqref{eq:classicalenergy2} is minimized all over five parameters, and we find the optimal skyrmion configuration on a stripe. 
The optimal values of parameters rapidly converge with increasing   $N_{Sk}$.

\subsection{The details of calculation}

Knowing the optimal parameters for each half plane, we calculate separately the band structure of SkX for each area. 
In our calculation with parameters $d$, $z_0$, $z_0'$, $\alpha$ and $\alpha'$, we consider the magnon states of SkX on the finite stripe 
and it means that each branch of the spectrum for given $q_{x}$ splits according to the number of skyrmions in the stripe, $N_{Sk}$.


Solving  the equation \eqref{eq:shr} we notice that the imposed periodicity of the stripe geometry allows us 
to use the plane-wave basis and represent the wave functions and potentials by Fourier series. 
The new elementary cell is elongated along the $y$-axis,  being $(d,\sqrt{3} d N_{Sk})$, so that the first Brillouin zone is given by a  rectangle squeezed in the same direction, and  reciprocal lattice vectors are : $\mathbf{b}_{1}=(2\pi/d,0)$ and $\mathbf{b}_{2}=(0,2\pi/\sqrt{3}dN_{Sk})$.

We seek a solution of the stationary equation \eqref{eq:BdGeq} in the following form:
\begin{equation}
\begin{pmatrix}
  \psi\\
  \bar{\psi}
\end{pmatrix} =
e^{iq_xx}\begin{pmatrix}
  u(\mathbf{r})\\
  v(\mathbf{r})
\end{pmatrix}=e^{iq_xx}\sum\limits_{\mathbf{Q}}e^{i\mathbf{Q}\mathbf{r}}
\begin{pmatrix}
  u_\mathbf{Q}\\
  v_\mathbf{Q}
\end{pmatrix}\,,
\label{eq:bloch}
\end{equation}
where $\mathbf{Q}=n\mathbf{b}_{1}+m\mathbf{b}_{2}$ with integers $n$ and $m$. The potentials in the Hamiltonian $\mathcal{H}$ expand in the same manner, $R(\mathbf{r}) = \sum_\mathbf{Q}R_{\mathbf{Q}}\exp{(i\mathbf{Q}\mathbf{r})}$, with $R$  replaced by scalar potentials $U$,$V$ or vector potential $\mathbf{A}$. 
We note that the application of Bloch's theorem \eqref{eq:bloch} is allowed by the average zero flux of the field intensity,  $\int d\mathbf{r}\, \mbox{curl }\mathbf{A} =0$.


Using these expressions in Eq.\  \eqref{eq:BdGeq}, we obtain a set of linear equations for the Fourier series coefficients of the wave function. The search for energies of normal modes is thus reduced to the diagonalization  of the Hamiltonian matrix.

The calculation of eigenmodes is hindered by two factors. First, the vector potential $\mathbf{A}$ has mild singularities at skyrmion centers, 
 and we employ a regularization procedure  to improve the convergence of the solution with increasing the size of the basis, see  \cite{Timofeev2022} for details. 

Another complication arises from the anisotropy in the grid of reciprocal lattice vectors. We saw above, that the reciprocal lattice vector along the $y$-axis is significantly shorter than its counterpart along the $x$-axis. To ensure accurate results, we included a larger number of $y$-direction vectors into our calculations, thus maintaining a uniform covering of higher-order harmonics in momentum space.

\subsection{Results for the spectrum}

\begin{figure*}[t]
\center{\includegraphics[width=0.99\linewidth]{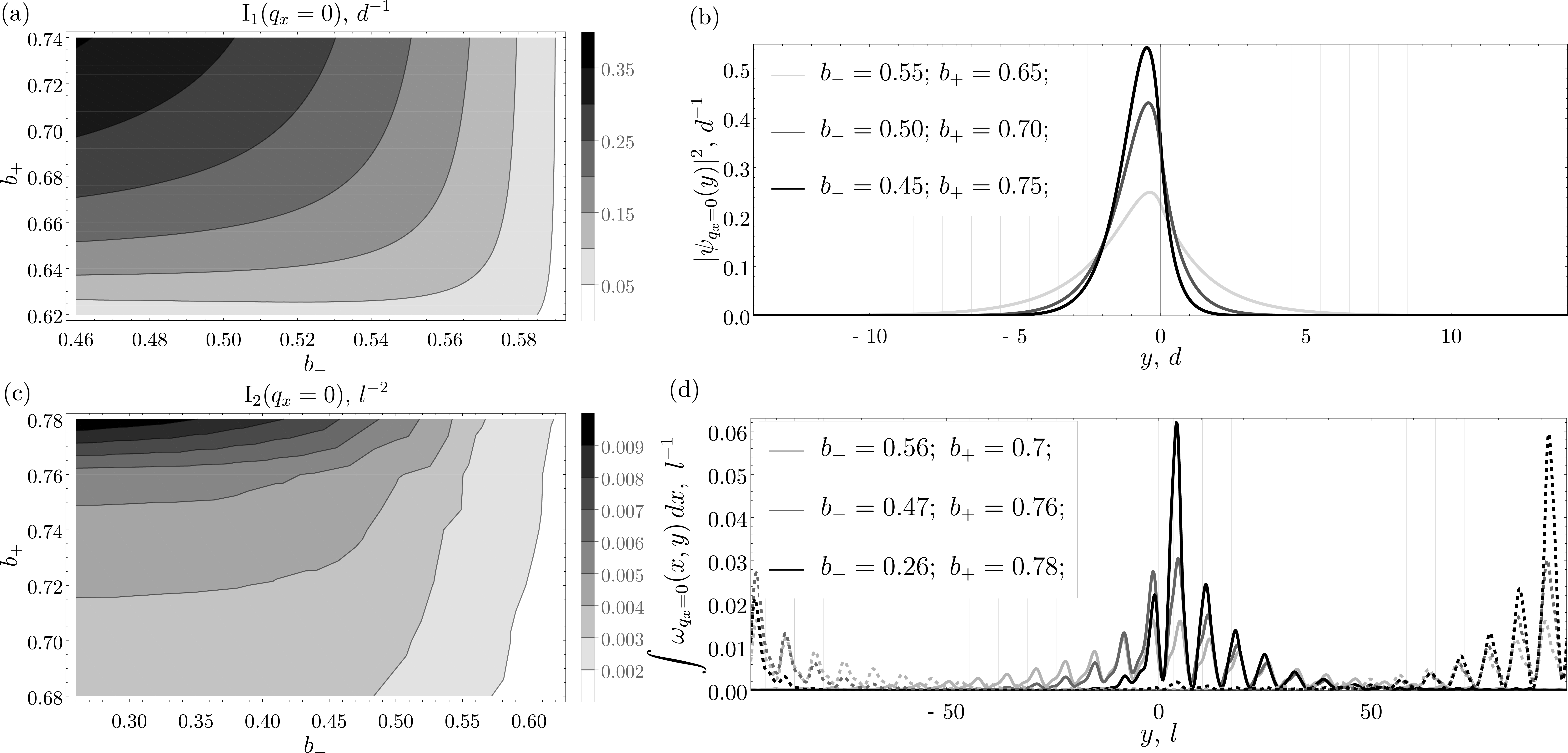}}
\caption{
\label{fig:numerics}
  The inverse participation ratio and the shape of the wave function for the edge states at the interface. 
Panels (a) and (b) correspond to the reduced model, Eq.\eqref{eq:redham}, showing  (a)  the dependence of IPR Eq.\eqref{eq:ipr1} on fields $b_-$ and $b_+$, (b) the modulus square of wave function for three combinations of the fields, $b_-$ and $b_+$, here shades of gray for curves correspond to shades of gray in the panel (a). Panels (c) and (d) show the same quantities as in panels (a) and (b), but calculated in the full model Eq.\eqref{eq:BdGeq}. The panel (c) shows IPR value, Eq.\eqref{eq:ipr2}, for different values of $b_-$ and $b_+$, and the panel (d) depicts the profile of densities $\int dx(|u(\mathbf{r})|^2-|v(\mathbf{r})|^2)$. Here dashed lines correspond to spurious states localized 
away from the interface, due to imposed cyclic boundary conditions, these states would disappear in the infinite sample geometry. Thin vertical lines correspond to positions of skyrmions' centers, see Fig.\ \ref{fig:statconfig}(b). 
}

\end{figure*}

We calculated the band structure of elementary excitations in a SkX stripe under a nonuniform magnetic field. The results 
for three pairs of field values $(b_{-},b_{+})$ and $N_{Sk}=14$ are  presented in Fig.\ \ref{fig:numerics} (a)-(c).  
Due to dimensional quantization, the dispersion $\epsilon(q_{x},q_{y})$ for the infinite SkX is replaced by curves $\epsilon_{j}(q_{x})$ with $1\le j\le 2N_{Sk}$  for the finite stripe of length $2N_{Sk}$.

For better visualization we also show in the same graphs the bands $\epsilon(q_{x}, q_{y})$ with $q_{y}\in (0,\sqrt{3} \pi/d)$ calculated for uniform planes with $b= b_{-}$ (blue stripes) and $b= b_{+}$ (red stripes) 
These bands correspond to our  calculation of the band structure in SkX in the uniform magnetic field \cite{Timofeev2022}. 

 One can see a good overlap between the dispersion of magnons in case of nonuniform magnetic field and the dispersions of magnons at the uniform field values, except for one thing.  
Namely, one see in Fig.\ \ref{fig:dispers} the existence of  localized modes, whose dispersion lies within the gap between two bulk band dispersions.  Let us discuss these modes in more detail.

First, we see that the number of these modes is two, since our model setup is periodic in $y$-direction and there are  two interfaces between red and blue dots in Fig.\ \ref{fig:statconfig}. 
As one can see from Fig.\ref{fig:dispers} the resulting two intra-gap modes are symmetrical with opposite group velocities.  
Below we show that these two edge states are localized at the opposite ends of our stripe.
One might say, that  Fig.\ref{fig:dispers} shows a true edge state, with negative group velocity, and another one with positive group velocity, that would be localized at infinite distance in case of infinite plane. 
We note in passing that if  the region on the plane with greater value of the field, $b>b_{tt}$, would be produced by micro solenoid, then this region could be confined within the sample and in this case there would be only one edge state with circular flow of magnon current. 



Second, we observe that similarly to simplified model of Sec. \ref{sec:toptrans}, the edge states in Fig.\ref{fig:dispers} have localized character and we discuss it in the next section.

\section{Localized states
\label{sec:IPR}}

{ The degree of localization of interface modes can be characterized in terms of inverse participation ratio (IPR).  In quantum mechanics the square of wave function, $w(\mathbf{r})=|\psi(\mathbf{r})|^2$, gives a probability to find a particle at a position $\mathbf{r}$ in some region $\Omega$, with the normalization  $ \int_{\Omega}d\mathbf{r}\,w(\mathbf{r})=1$. 
 The  IPR quantity, $I$, is defined then as an integral of a square of this weight over this region, $ {I}=\int_{\Omega}  d\mathbf{r}\,w^2(\mathbf{r})$. We discuss two cases above: (i) reduced continuous model with a solution, Eq.\ \eqref{eq:half-planesol}, localized in one direction and (ii) full model with a quasi-periodic solution \eqref{eq:bloch} on a finite stripe and different orthogonality condition.  Hence we consider two different IPR definition for these cases. For the reduced model we define:
\begin{equation}
      {I}_{1}(q_x)=\int\limits_{-\infty}^{{\infty}} dy\,\left(\Psi^{\dagger}_{q_x}(\mathbf{r}) \Psi_{q_x}(\mathbf{r})\right) ^2 \,,
\label{eq:ipr1}
\end{equation}
with the normalization $\int\limits_{-\infty}^{{\infty}} dy\,\Psi^{\dagger}_{q_x}(\mathbf{r}) \Psi_{q_x}(\mathbf{r}) =1$.

In case of full model calculation we define the IPR as follows:
\begin{equation}
    {I}_2(q_x)=\int\limits_{\Omega} d\mathbf{r}\,w^2(\mathbf{r}) \,,
\label{eq:ipr2}
\end{equation}
with the weight  $w(\mathbf{r}) = |u(\mathbf{r})|^2-|v(\mathbf{r})|^2$,  and $u(\mathbf{r})$ and $v(\mathbf{r})$   components of spinor \eqref{eq:bloch}, $\Omega$ corresponds to the stripe in Fig.\ref{fig:statconfig}. The normalization in this case is $ \int_{\Omega} \mathbf{r}\,w(\mathbf{r}) =1$, see the note after Eq.\ \eqref{eq:shr}. 
In this case of  2D stripe the IPR is measured in units of $l^{-2}$. 


We calculated the IPR criterion for the intra-gap edge state with  $q_x=0$ and for the range  of pairs $(b_{-},b_{+})$ with 
 $b_{-} < b_{tt}< b_{+}$, and show the results in Fig.\ref{fig:numerics}(a) for the reduced model and in Fig.\ref{fig:numerics}(c) for the full one}. The darker region corresponds to more localized character of the state, which happens when both $b_{-}$ and  $b_{+}$ are far away from  $b_{tt}$.  
The edge state become more delocalized in case (i) either $b_{+}$, or $b_{-}$, or both come close to $b_{tt}$ and (ii) the bulk flat band crosses the energy value of our interest, with the hybridization between edge and flat band states.


We illustrate the localized character of the edge state    in Fig.\ \ref{fig:numerics}(d) for three pairs of $(b_{-},\,b_{+})$.  
The localization  is clearly visible here, while 
the peaks in values of $w(\mathbf{r}) = |u(\mathbf{r})|^2-|v(\mathbf{r})|^2$  correspond to positions of skyrmions in Fig. \ref{fig:statconfig}.  
It is more instructive to express the degree of localization in terms of interskyrmion distance. Multiplying the calculated IPR by the area of a triangular SkX's unit cell, $N_{loc}=( {I}_2\times \sqrt3d^2/2)^{-1}$, gives us the number of skyrmions over which the edge state is localized. This number  is given in Table~ \ref{tab:IPR_numb} for  three  pairs, $(b_{-},\,b_{+})$, and is in accordance with Fig.\ \ref{fig:numerics}(d).



\begin{table}[t]
\caption{The degree of localization expressed in terms of  IPR and as the number of  SkX unit cells, over which the edge state is localized. 
\label{tab:IPR_numb} }  
\centering
\begin{tabular}{|c||c|c|c|}
\hline
$(b_{-},\,b_{+})$ & $(0.26,\,0.78)$ & $(0.47,\,0.76)$ & $(0.56,\,0.7)$ \\
\hline
$ {I}_2(q_x=0),\,l^{-2}$ & 0.0099 & 0.0045 & 0.0025 \\
\hline
$N_{loc}$ & 1.86 & 4.15 & 7.54 \\
\hline
\end{tabular}
\end{table}

%
 
\section{Conclusion}
\label{sec:conc}

We studied the magnon spectrum of thin ferromagnetic films with DMI placed in a spatially nonuniform magnetic field. We considered a situation when one half of the sample is placed in a field below the  topological transition point, $b < b_{tt}$, and another half has a field above this point, $b > b_{tt}$. We demonstrated earlier that the dispersion of low-energy magnons -- corresponding to breathing  and counterclockwise  modes near the Brillouin zone center -- can be effectively described by the extended Dirac equation, including terms quadratic in momentum. Such simplified modeling gives us a qualitative description of the edge states emerging at the interface between the regions with different magnetic field strength.

Within the stereographic projection approach, we further performed a full-scale calculation of the elementary excitation spectrum in an elongated strip geometry under spatially nonuniform magnetic fields. Staying in the harmonic approximation (equivalent to linear spin-wave theory), we  confirmed the appearance of edge states localized near the interface between two distinct field regions. The appearing edges states are chiral in  nature,  and their group velocity  along the interface is determined by the cross product of the normal vector to the film and the field gradient. It means that the heat flow transferred by the edge states can be controlled by the external magnetic field configuration.

We demonstrate that these edge states remain localized at the interface within a few lattice periods of SkX
 across a wide range of magnetic fields. 
 In real materials, this spatial scale may correspond to several hundred nanometers. Owing to this unique property, SkX under nonuniform magnetic fields hold significant promise for magnonics applications, particularly for engineering field-controllable quasi-one-dimensional magnon waveguides.
We anticipate that the results presented in this work will stimulate the search for candidate compounds and systems where such chiral magnon localization phenomena could be realized.

\begin{acknowledgments} 
V.T. thank  R.A. Niyazov for useful discussions and provided computing resources. The work was supported by the Russian Science Foundation, Grant No. 24-72-00083. The work of authors is  partially supported by the Foundation for the Advancement of Theoretical Physics BASIS.
\end{acknowledgments}

\bibliography{skyrmionbib}
\end{document}